\documentclass[seceq]{ptptex}

\usepackage{graphicx}

\title{
Analytical estimates of the locations of phase transition points in the ground state for the  bimodal Ising spin glass model in two dimensions%
}

\author{
Chiaki Yamaguchi%
}

\inst{
Kosugichou 1-359, Kawasaki 211-0063, Japan
}

\abst{
 We analytically estimate the locations of phase transition points
 in the ground state for the $\pm J$ random bond Ising model
 with asymmetric bond distributions on the square lattice.
 We propose and study the percolation transitions
 for two types of bond shared by two non-frustrated plaquettes.
 The present method indirectly treats the sizes of clusters of correlated spins
 for the ferromagnetic and spin glass orders. 
 We find two transition points.
 The first transition point is the phase transition point for the ferromagnetic order, and the location is obtained as $p_c^{(1)} \approx 0.895 \, 399 \, 54$ as the solution of $[p^2 + 3 (1-p)^2 ]^2 \, p^3 - \frac{1}{2} = 0$.
 The second transition point is the phase transition point for the spin glass order, and the location is obtained as $p_c^{(2)} = \frac{1}{4} [2 + \sqrt{2 (\sqrt{5} - 1)}] \approx 0.893 \, 075 \, 69$.
 Here, $p$ is the ferromagnetic bond concentration,
 and $1 - p$ is the antiferromagnetic bond concentration.
 The obtained locations are reasonably close to the previously estimated locations.
 This study suggests the presence of
 the intermediate phase between $p_c^{(1)}$ and $p_c^{(2)}$; however,
 since the present method produces remarkable values but 
 has no mathematical proof for accuracy yet, no conclusions are drawn
 in this article about the presence of the intermediate phase. 
}

\begin{document}
\maketitle

\section{Introduction} \label{sec:1}

The establishment of reliable theories of spin glasses has been one of the
most challenging problems in statistical physics for years\cite{EA1975, KR, N2001, MPV, T}.
In this article, our main interests are
 to make the determination of the structure of the phase diagram for spin glasses,
 and to clarify the properties of the phases.

\begin{figure}[t]
\begin{center}
\includegraphics[width=0.30\linewidth]{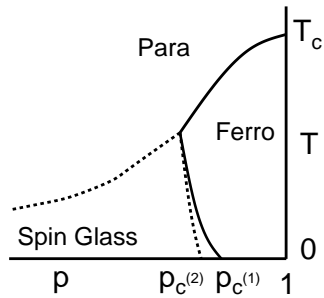}
\end{center}
\caption{
 A schematic phase diagram for the $\pm J$ random bond Ising model
 on the square lattice. The relationship between
 the ferromagnetic bond concentration $p$ and the temperature $T$ is shown,
 where $1 - p$ is the antiferromagnetic bond concentration.
 $p_c^{(1)}$ and $p_c^{(2)}$ stand for phase transition points at zero temperature. 
 $T_c$ stands for the critical temperature of
 the ferromagnetic model.
 Para, Ferro, and Spin Glass stand for the paramagnetic,
 ferromagnetic and spin glass phases, respectively.
 The phase between  $p_c^{(1)}$ and $p_c^{(2)}$
 is the intermediate phase.
\label{fig:figure1}
}
\end{figure}
 Figure~\ref{fig:figure1} shows
 a schematic phase diagram for the $\pm J$ random bond Ising model
 on the square lattice.
 In this article, we do not mention the problem\cite{KA2000}
 of the presence or absence of the spin glass phase at a finite temperature
 for the square lattice.
 In this article, we mention the locations of
 the phase transition points $p_c^{(1)}$ and $p_c^{(2)}$ at zero temperature for the square lattice
 and mention the intermediate phase between $p_c^{(1)}$ and $p_c^{(2)}$.
 The  exact locations of $p_c^{(1)}$ and $p_c^{(2)}$ have not been
 analytically obtained.
 One of the aims of this article is to analytically estimate
 the two locations.
 
We examine the bond configurations for the exchange interaction.
 We propose and study the percolation transitions
 for two types of bond shared by two non-frustrated plaquettes.
 The present method indirectly treats the sizes of clusters of correlated spins
 for the ferromagnetic and spin glass orders. 
 A similar scheme has already been proposed by Sadiq {\it et al.}
 \cite{SKWB1981}.
 Sadiq {\it et al.} investigated 
 the plaquette percolation for frustrated
 and non-frustrated plaquettes by computer simulation,
 and argued a connection with statistical properties at zero temperature.
 We show in this article
 that the present percolation and the plaquette percolation\cite{SKWB1981}
 yield different results.
 Recently, Miyazaki\cite{M2013} proposed
 an analytical method that estimates
 the location of the phase transition point at zero temperature
 by analyzing the number of frustrated plaquettes.
 Miyazaki's method estimates one location for the square lattice\cite{M2013}.
 On the other hand, e.g. by computing the domain wall energy,
 it is possible to numerically estimate two locations
 for the square lattice\cite{KR1997},
 although there is a problem with the agreement or disagreement
 of the two locations\cite{KR1997}.
 In this study, the two locations of the phase transition points are analytically estimated.

Maynard and Rammal \cite{MR1982} proposed that a highly correlated state,
 called the random antiphase state (RAS), is present near the ferromagnetic phase
 in the low temperature region, but Morgenstern \cite{M1982} claimed that
 a transfer-matrix calculation indicates the absence of the RAS.
 Kawashima and Rieger \cite{KR1997} investigated
 the ground state by an exact numerical polynomial method,
 estimated the two locations of the phase transition points,
 and concluded that the intermediate phase is absent in the ground state.
 The locations estimated in this study are
 remarkably close to the locations estimated by Kawashima and Rieger.
 On the other hand, the present study suggests the presence of the intermediate phase.
 We discuss the properties of phases
 by using the present bonds, and give another interpretation
 of previous numerical results.
 One of the aims of this article is to clarify the properties of
 the phases at zero temperature.

In order to calculate the locations,
 the approximate approach, regarding the present bonds as random bonds,
 is applied.
 By comparing the obtained locations with the previously estimated locations,
 we show the validity of the present results.
 Lower bounds of the locations with no approximation
 by using the present bonds are also obtained.
 Most of the present results are results for the square lattice.
 Application of the present scheme
 to the model on a simple cubic lattice is also mentioned.

This article is organized as follows.
 We explain the present model and the concepts
 of frustration and bond percolation in Sect.\ref{sec:2}.
 These concepts are used in this study.
 The analytical estimates of the locations of the phase transition points
 are given in Sect.\ref{sec:3}.
 The properties of phases are discussed in Sect.\ref{sec:4}.
 In Sect.\ref{sec:5}, another interpretation of
 previous numerical results is given.
 The concluding remarks for this article are in Sect.\ref{sec:6}.

\section{The model and the concepts of frustration and bond percolation} \label{sec:2}

The Hamiltonian ${\cal H}$ for the $\pm J$ random bond Ising model
 is given by  \cite{KR, N2001, MPV, T, EA1975}
\begin{equation}
 {\cal H} = - \sum_{\langle i j \rangle} J_{i j} S_i S_j \, ,
\end{equation}
 where $\langle i j \rangle$ denotes nearest-neighbor pairs, $S_i$ is
 a state of the spin at site $i$, and $S_i = \pm 1$.
 $J_{i j}$ is a strength of the exchange interaction between the spins at sites $i$ and $j$,
 and is a quenched variable.
 The value of $J_{i j}$ is given with
  the distribution $P (J_{i j})$, and $P (J_{i j})$ is given by
\begin{equation}
 P (J_{i j})
 = p \, \delta_{J_{i j}, 1} + (1 - p) \, \delta_{J_{i j}, - 1} \, ,
 \label{eq:PpmJJij}
\end{equation}
 where  $\delta$ is the Kronecker delta.
 When $J_{i j} = 1$, the interaction is also called
 the ferromagnetic bond.
 When $J_{i j} = - 1$, the interaction is also called
 the antiferromagnetic bond.
 A schematic phase diagram for this model on the square lattice  
 is shown in Fig.~\ref{fig:figure1}.

\begin{figure}[t]
\begin{center}
\includegraphics[width=0.30\linewidth]{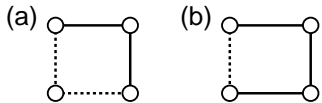}
\end{center}
\caption{
 Examples of plaquettes. The solid lines represent the ferromagnetic bonds, and the dotted lines represent the antiferromagnetic bonds. (a) A non-frustrated plaquette. (b) A frustrated plaquette.
\label{fig:figure2}
}
\end{figure}
 Figure~\ref{fig:figure2} shows examples of plaquettes.
 In Fig.~\ref{fig:figure2} (a), there are ways of choosing
 the orientations of the spins on the four sites without frustration. 
 On the other hand, in Fig.~\ref{fig:figure2} (b), there is no way of choosing
 the orientations of the spins on the four sites without frustrating at least one bond. 
 This frustration effect is measured with the frustration function $\Phi$, given by \cite{T}
\begin{equation}
 \Phi = \prod_{\{ c \} } J_{i j} \, ,
\end{equation}
 where $\{ c \}$ is a set of closed contours. 
 When there is a frustrated plaquette, $\Phi$ gives $- 1$ and gives $1$ otherwise.
 For example, $\Phi$ of Fig.~\ref{fig:figure2} (a) gives $1$, and $\Phi$ of Fig.~\ref{fig:figure2} (b) gives $- 1$.

In the bond percolation, bonds for generating clusters are
 placed on the edges of the lattice, and one of the clusters is percolated
 at the threshold fraction $P_c$ \cite{K1973}.
 For example, in Fig.~\ref{fig:figure3},
 the clusters for the p1 bonds are not percolated,
 and  the cluster for the p2 bonds is percolated.
 The details of Fig.~\ref{fig:figure3} and the p1 and p2 bonds are given in Sect.\ref{sec:3}.

\section{Analytical estimates of the locations of the phase transition points} \label{sec:3}

We analytically estimate the locations of the phase transition points
 at zero temperature for the square lattice.
 The first step is to extract and analyze the non-frustration network,
 since it is considered that
 the local energy on the non-frustration network tends to be lower than 
 that on the frustration network.
 In addition, we consider effective bonds that describe phase transitions.
 Thus, it is speculated that
 such a bond is shared by two non-frustrated plaquettes. 
 This bond has two types: we call them p1 and p2 bonds.
 We call the ferromagnetic bond,
 which is shared by two non-frustrated plaquettes, the p1 bond.
 We call the bond that includes ferromagnetic and antiferromagnetic bonds and
 that is shared by two non-frustrated plaquettes the p2 bond.

We expect that the sizes of the clusters for the p1 bonds correspond to
 the sizes of clusters of correlated spins for the ferromagnetic order and
 expect that the sizes of the clusters for the p2 bonds
 correspond to the sizes of clusters of correlated spins for the spin glass order. 
 The correlation for the ferromagnetic order at sites $i$ and $j$
 is given by $[ \langle S_i S_j \rangle_\beta ]_J$, where $\langle \, \rangle_\beta$
 denotes the thermal average and
 $[ \,]_J$ denotes the configurational average for the exchange interactions.
  The correlation for the spin glass order at sites $i$ and $j$
 is given by $[ \langle S_i S_j \rangle^2_\beta ]_J$.

 The present method has no mathematical proof for accuracy yet; on the other hand,
 from the definitions of the present bonds in this section and the arguments
 for phases by using the present bonds in Sect.\ref{sec:4}, connections between
 the sizes of clusters of correlated spins for the orders and
 the sizes of the clusters for the present bonds are inferred.
 The degrees of these connections are related to
 the accuracies of the present results. 
 The validity of the present bonds is shown from
 the percolation thresholds in this section.

\begin{figure}[t]
\begin{center}
\includegraphics[width=0.45\linewidth]{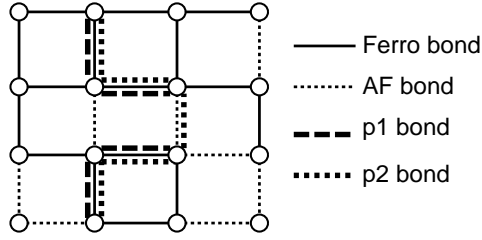}
\end{center}
\caption{
 Examples of four bonds.
\label{fig:figure3}
}
\end{figure}
 Figure~\ref{fig:figure3} shows  examples of four bonds.
 The solid lines represent the ferromagnetic bonds,
 the dotted lines represent the antiferromagnetic bonds,
 the dashed lines represent the p1 bonds, and
 the short dashed lines represent the p2 bonds.
 As shown in the figure,
 the p1 and p2 bonds are placed on the edges of the lattice.
 We analytically estimate the percolation thresholds
 for the p1 and p2 bonds on the square lattice.

\begin{figure}[t]
\begin{center}
\includegraphics[width=0.45\linewidth]{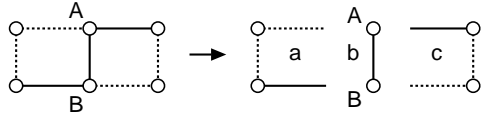}
\end{center}
\caption{
 Two plaquettes that are separated into three parts
 $a$, $b$, and $c$ for analytical calculation.
 The p1 and p2 bonds are placed on the edge $AB$.
\label{fig:figure4}
}
\end{figure}
 Figure~\ref{fig:figure4} shows two plaquettes that are separated into three parts $a$, $b$, and $c$
 for analytical calculation.
 The p1 and p2 bonds are placed on the edge $AB$ of Fig.~\ref{fig:figure4}.
 By generalizing these placements, we calculate
 the probabilities for placing the present bonds on an edge. 
 Part $a$ of  Fig.~\ref{fig:figure4} consists of three bonds.
 The probability $P^{(a)}_{+}$ that
 the frustration function $\Phi^{(a)}$ for part $a$ gives $1$ is given by
\begin{equation}
 P^{(a)}_{+} = \left( \begin{array}{c} 3 \\ 0 \end{array} \right) p^3
 +  \left( \begin{array}{c} 3 \\ 2 \end{array} \right) p (1 - p)^2 
 = p^3 + 3 p (1 - p)^2 \, ,
\end{equation}
 where
$$
 \left( \begin{array}{c} l \\ m \end{array} \right) \equiv
 \frac{l !}{m ! (l - m) !} \, . 
$$
 The probability $P^{(a)}_{-}$ that
 the frustration function $\Phi^{(a)}$ gives $-1$ is given by
\begin{equation}
P^{(a)}_{-} = \left( \begin{array}{c} 3 \\ 1 \end{array} \right) p^2 (1 - p)
 +  \left( \begin{array}{c} 3 \\ 3 \end{array} \right) (1 - p)^3
 = 3 p^2 (1 - p) + (1 - p)^3  \, .
\end{equation}
 Part $b$ of  Fig.~\ref{fig:figure4} consists of a bond.
 Similarly, $P^{(b)}_{+}$ and $P^{(b)}_{-}$ are respectively given by
\begin{equation}
 P^{(b)}_{+} = p \quad {\rm and} \quad P^{(b)}_{-} = 1 - p \, .
\end{equation}
 In addition, $P^{(c)}_{+}$ and $P^{(c)}_{-}$ are respectively
\begin{equation}
 P^{(c)}_{+} = P^{(a)}_{+} \quad {\rm and} \quad P^{(c)}_{-} = P^{(a)}_{-} \, .
\end{equation}

 The probability $P^{({\rm p1 \, bond})}$ for placing the p1 bond on an edge is obtained as
\begin{equation}
 P^{({\rm p1 \, bond})} = P^{(a)}_{+} \, P^{(b)}_{+} \, P^{(c)}_{+} 
 = p^3 [p^2 + 3 (1-p)^2 ]^2 \, . \label{eq:eq1}
\end{equation}
 At the percolation transition point, there is a relation:
\begin{equation}
 P^{({\rm p1 \, bond})} = P^{({\rm p1 \, bond})}_c \, , \label{eq:eq2}
\end{equation}
 where $P^{({\rm p1 \, bond})}_c $ is the percolation threshold
 for the p1 bond.
 Calculation of $P^{({\rm p1 \, bond})}_c $ is not trivial;
 thus, here we approximate the $P^{({\rm p1 \, bond})}_c $ 
 by regarding 
 the p1 bond as a random bond.
 The percolation threshold of the random bond for the square lattice
 is $\frac{1}{2}$ \cite{K1973};
 therefore, we obtain $P^{({\rm p1 \, bond})}_c  \approx \frac{1}{2}$. 
 Therefore, by solving the equation $p^3 [p^2 + 3 (1-p)^2 ]^2 = \frac{1}{2}$,
 we obtain the location of the phase transition point $p_c^{(1)}$ for
 the ferromagnetic order as 
$$
 p_c^{(1)} \approx 0.895 \, 399 \, 54 \, .
$$ 
The obtained location is consistent with 
previous estimates such as 0.8955(11) \cite{JP2012},
 0.896(1) \cite{KR1997}, 0.88(2) \cite{MK1980},
 0.89(2) \cite{BMRU1982} and 0.89(1) \cite{ON1987a, ON1987b}, and
 the obtained location is smaller than but
 marginally consistent with 0.8969 \cite{M2013},
 while the obtained location is inconsistent with
 0.897(1) \cite{AH2004}, 0.8969(1) \cite{WHP2003} and 0.885(1) \cite{O1990}.

The probability $P^{({\rm p2 \, bond})}$ for placing the p2 bond on an edge is obtained as
\begin{eqnarray}
 P^{({\rm p2 \, bond})} &=&  P^{(a)}_{+} \, P^{(b)}_{+} \, P^{(c)}_{+} +  P^{(a)}_{-} \, P^{(b)}_{-} \, P^{(c)}_{-} \nonumber \\
 &=&  p^3 [p^2 + 3 (1-p)^2 ]^2+  (1 - p)^3  [ 3 p^2 + (1 - p)^2 ]^2 \, . \label{eq:eq3}
\end{eqnarray}
 At the percolation transition point, there is a relation:
\begin{equation}
 P^{({\rm p2 \, bond})} = P^{({\rm p2 \, bond})}_c \, , \label{eq:eq4}
\end{equation}
 where $P^{({\rm p2 \, bond})}_c $ is the percolation threshold
 for the p2 bond.
 Calculation of $P^{({\rm p2 \, bond})}_c $ is not trivial;
 thus, here we approximate the $P^{({\rm p2 \, bond})}_c $ 
 by regarding 
 the p2 bond as a random bond.
 The percolation threshold of the random bond for the square lattice
 is $\frac{1}{2}$ \cite{K1973};
 therefore, we obtain $P^{({\rm p2 \, bond})}_c  \approx \frac{1}{2}$.
 Therefore, by solving the equation
 $p^3 [p^2 + 3 (1-p)^2 ]^2 + (1 - p)^3 [ 3 p^2 + (1 - p)^2 ]^2 
 = \frac{1}{2}$,
 we obtain the location of the phase transition point $p_c^{(2)}$
 for the spin glass order as 
$$
 p_c^{(2)} = \frac{1}{4} [2 + \sqrt{2 (\sqrt{5} - 1)}] \approx 0.893 \, 075 \, 69 \, .
$$ 
The obtained location is consistent with 
 a previous estimate such as 0.894(2) \cite{KR1997}, and
 the obtained location is larger than but marginally consistent with 
 0.85 \cite{BMRU1982} and 0.870 \cite{O1995},
 while the obtained location is inconsistent with 0.896(2) \cite{AH2004}, 
 0.86(2) \cite{SKWB1981} and 0.854(2) \cite{O1990}.

As described above, the obtained locations of $p_c^{(1)}$ and $p_c^{(2)}$ are
 reasonably close to the previously estimated locations, although
 the approximate approach by regarding the present bonds as random bonds
 is applied.
 Thus, it is considered that this approximate approach
 is valid in this study.
 We derive the lower bounds of the locations
 with no approximation by using the present bonds.
 The present bonds are slightly more complicated than the random bonds,
 so the exact values of $P^{({\rm p1 \, bond})}_c$
 and $P^{({\rm p2 \, bond})}_c$ can be slightly larger than $\frac{1}{2}$:
 $P^{({\rm p1 \, bond})}_c > \frac{1}{2}$ and
 $P^{({\rm p2 \, bond})}_c > \frac{1}{2}$.
 Then, by using Eqs.~(\ref{eq:eq1}), (\ref{eq:eq2}), (\ref{eq:eq3}), and (\ref{eq:eq4}),
 we obtain the lower bounds for $p_c^{(1)}$ and  $p_c^{(2)}$
 with no approximation as $0.895 \, 3995\cdots < p_c^{(1)} ({\rm no \, approximation})$
 and $0.893 \, 0756\cdots < p_c^{(2)} ({\rm no \, approximation})$.
 From one of these inequalities, it is shown that
 the present percolation and the plaquette percolation\cite{SKWB1981}
 yield different results, since the plaquette percolation
 gives $p_c^{(2)} = 0.86(2)$ \cite{SKWB1981}.

We consider application of the present scheme
 to the model on a simple cubic lattice.
 We consider a bond shared by four plaquettes.
 It is considered that
 the $q_4$ bond shared by four non-frustrated plaquettes
 is effective, but it is supposed that
 the $q_3$ bond shared
 by three non-frustrated plaquettes and a frustrated plaquette
 is also effective.
 The percolation threshold of the random bond for the simple cubic lattice
 is about 0.248 \, 81\cite{WZZGD2013, DH2004, LZ1998}.
 Therefore, the equation for the ferromagnetic phase transition
 is given by
 $4 x p^4 (1- p) [ p^2 + 3 (1 - p)^2 ]^3  [ 3 p^2 + (1 - p)^2 ]
  + p^5 [ p^2 + 3 (1 - p)^2 ]^4 = 0.248 \, 81$, where $0 \le x \le 1$.
 In this equation, the effect for the ferromagnetic $q_4$ bond is fully counted.
 When $x = 0$, the effect for the ferromagnetic $q_3$ bond is not counted, and
 the result is given by $p^{(x = 0)}_c \approx 0.884 \, 95$.
 When $x = 1$, the effect for the ferromagnetic $q_3$ bond is fully counted, and
 the result is given by $p^{(x = 1)}_c \approx 0.694 \, 79$. 
 The previous numerical estimates for $p_c$ are
 0.7747(7) \cite{TK2011} and 0.778(5) \cite{H1999}.
 Compared to the previous numerical estimates,
 the value of $x$ takes a value for $0 < x < 1$,
 and therefore the ferromagnetic $q_3$ bond is
 also related to the ferromagnetic phase transition.

\section{The properties of phases} \label{sec:4}

We now discuss the properties of phases at zero temperature for the square lattice.
 We call the antiferromagnetic bond,
 which is shared by two non-frustrated plaquettes, the p3 bond.
 The probability $P^{({\rm p3 \, bond})}$ for placing the p3 bond on an edge is obtained as
\begin{equation}
 P^{({\rm p3 \, bond})} = P^{(a)}_{-} \, P^{(b)}_{-} \, P^{(c)}_{-}
 = (1 - p)^3 [ 3 p^2 + (1 - p)^2 ]^2 \, .
\end{equation}
 One can see that $P^{({\rm p3 \, bond})} > 0$ for $\frac{1}{2} \le p < 1$.
 Therefore, non-frustration networks always include the p3 bonds
 for $\frac{1}{2} \le p < 1$ if the system size is large enough.
 There is no percolation transition for clusters that consist only of the p3 bonds
 when $\frac{1}{2} \le p \le 1$.
 The p3 bond works as the antiferromagnetic interaction
 and breaks the ferromagnetism.
 The p1 and p2 bonds are defined in Sect.\ref{sec:3}.

We now discuss the ferromagnetic phase.
 The ferromagnetic correlation length of spins
 on the percolated p1 cluster in the ground state
 is equal to the system length.
 Then, the ferromagnetic phase appears.
 Next, we discuss the intermediate phase.

\begin{figure}[t]
\begin{center}
\includegraphics[width=0.25\linewidth]{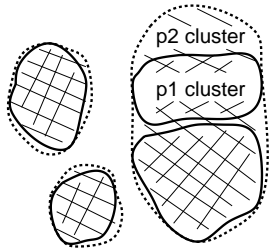}
\end{center}
\caption{
 A relationship for the p1 and p2 clusters.
\label{fig:figure5}
}
\end{figure}
 Figure~\ref{fig:figure5} shows
 a relationship for the p1 and p2 clusters.
 As shown in Fig.~\ref{fig:figure5},
 the sizes of the p2 clusters are always larger than those of the p1 clusters,
 since the p2 bond includes the p1 and p3 bonds.
 The intermediate phase appears
 when the p1 cluster is not percolated and the p2 cluster is percolated,
 as illustrated in Fig.~\ref{fig:figure3}.
 If the p1 cluster is not percolated, 
 then the p3 bonds break the ferromagnetism, and
 the ferromagnetism does not appear.
 In the intermediate phase, since the p2 cluster is percolated,
 the non-frustration network is percolated, and
 therefore the spins are in highly correlated states.
 Highly correlated states are a feature of the RAS.
 Thus, the nature of the intermediate phase is the RAS.

 According to the locations obtained in Sect.\ref{sec:3},
 the width of the intermediate phase is estimated
 as
$$
 p_c^{(1)} - p_c^{(2)} \approx 0.002 \, 323 \, 85 \, .
$$
 Thus, the width is very narrow.

 The probability $P_O$ of the occurrence of the antiferromagnetic bonds
 on the non-frustration network is obtained as
 $P_O = \frac{(1-p)^3 [3 p^2 + (1-p)^2 ]^2}{p^3 [ p^2 + 3 (1-p)^2 ]^2 + (1-p)^3 [3 p^2 + (1-p)^2 ]^2}$ by solving $P_O = P^{({\rm p3 \, bond})} / P^{({\rm p2 \, bond})}$.
 By using the values at $p = p_c^{(1)}$ and $p = p_c^{(2)}$,
 the range of $P_O$ for the intermediate phase
 is estimated as $$0.013 \, 18 < P_O < 0.014 \, 14 \, .$$
 Therefore, the antiferromagnetic bonds contribute to
 the non-frustration network at low probability.

 We also estimate the required system length $L_R$ for investigating
 the intermediate phase.
 In order to flip a cluster of spins in one dimension
 by using the p3 bonds,
 at least two p3 bonds are needed if the periodic boundary condition is implemented.
 Therefore, the required system length $L_R$ is given by $L_R > \frac{2}{P_O}$.
 By considering the range of the $P_O$ for the intermediate phase,
 the length is obtained as $$L_R > 151 \, .$$
 This inequality limits the system lengths that should be investigated.
 If the system length $L$ is shorter than $151$,
 the RAS does not appear because of the effect of finite system size,
 and the spin states indicate the ferromagnetic phase.
 If $L > 151$, the RAS appears.
 The required system sizes for investigating
 the intermediate phase are large.

We now discuss the spin glass phase.
 In the spin glass phase,
 the p1 and p2 clusters are not percolated, and
 thus the non-frustration network is not percolated.
 Therefore, the spin glass phase is different from the other phases, and
 the spins are not in highly correlated states.

\section{Another interpretation of previous numerical results} \label{sec:5}

We now give another interpretation of the previous numerical results.
 As described in Sect.\ref{sec:1}, Kawashima and Rieger \cite{KR1997} 
 concluded that the intermediate phase is absent 
 by using an exact numerical method.
 According to Kawashima and Rieger\cite{KR1997},
 the location of $p_c^{(1)}$ is estimated as 0.896(1),
 and the location of $p_c^{(2)}$ is estimated as 0.894(2).
 The difference is estimated as 0.002(2).
 According to Amoruso and Hartmann \cite{AH2004} by using an exact numerical method,
 the location of $p_c^{(1)}$ is estimated as 0.897(1),
 and the location of $p_c^{(2)}$ is estimated as 0.896(2).
 The difference is estimated as 0.001(2).
 On the other hand, in this study, as described in Sect.\ref{sec:3},
 the location of $p_c^{(1)}$ is estimated as about 0.8954,
 and the location of $p_c^{(2)}$ is estimated as about 0.8931.
 The difference is estimated as about 0.0023, as also described in Sect.\ref{sec:4}.
 Therefore, the difference for the present results
 is consistent with the differences for the previous results.
 If the intermediate phase is absent, the difference gives zero.
 Therefore, both interpretations of the presence and the absence
 are available when the previous results and the present result are compared.
 Thus, the precisions of the previous results are not sufficient for determining the phase.

 In addition, the required system length $L_R$ for investigating
 the intermediate phase is estimated as $L_R > 151$, as described in Sect.\ref{sec:4}.
 The largest system length $L_{\rm Max}$ investigated by Kawashima and Rieger \cite{KR1997}
 is $L_{\rm Max} = 32$, and  the length investigated by Amoruso and Hartmann \cite{AH2004}
 is $L_{\rm Max} = 700$.
 However, in the result by Amoruso and Hartmann,
 the data for $L < 151$ of the system length $L$
 were also used for estimating the locations of the phase transition points.
 Specifically, the data for $L = 80, 100, 120$, and $140$ were used
 in addition to the data for $L > 151$.
 According to the estimated required system length,
 the data for $L < 151$ should not be used.
 Thus, the system lengths investigated in the previous results
 are not sufficient for determining the phase.

\section{Concluding remarks} \label{sec:6}

We analytically estimated the locations of phase transition points
 in the ground state for the $\pm J$ random bond Ising model
 with asymmetric bond distributions on the square lattice.
 By comparing the obtained locations of transition points
 with the previously estimated locations,
 we showed the validity of the present results.
 We confirmed that 
 the present percolation transitions are
 very effectively related to the phase transitions at zero temperature.
 This indicates that the sizes of the clusters
 for the present bonds correspond to the sizes of clusters of correlated spins
 for the ferromagnetic and spin glass orders.
 
We also mentioned application of the present scheme
 to the model on the simple cubic lattice.
 It seems that the simple cubic lattice case is more complicated
 than the square lattice case.

We also discussed
 the properties of phases at zero temperature for the square lattice.
 The present study suggests that the intermediate phase
 between the ferromagnetic and spin glass phases
 is present and also suggests
 that the nature of the intermediate phase is the random antiphase state.
 Our arguments for the presence of the intermediate phase are summarized
 as follows: the antiferromagnetic bonds usually break
 the non-frustration network, but at low probability
 the antiferromagnetic bonds contribute to the non-frustration network,
 so that the intermediate phase is present
 with a very narrow width.

 Also, it was argued that
 the precisions of previous numerical results are not sufficient for determining the phase
 when the previous numerical results and the present result are compared.
 In addition, the required system length for investigating
 the intermediate phase was estimated and, by using the estimated required system length,
 it was argued that the system lengths investigated in the previous results
 are not sufficient for determining the phase.

 No conclusions, however, are drawn on
 the presence of the intermediate phase in this article,
 since the present method produces remarkable values but
 has no mathematical proof for accuracy yet. 
 More investigations with exact numerical methods are required to clarify
 the accuracies of the present percolations and
 the presence or absence of the intermediate phase.
 These investigations are tasks for the future.

In this study,
 the first step was to extract and analyze the non-frustration network,
 but this step contradicts the description of Ref.~\citen{T}, i.e.,
 ``Given some configuration of positive and negative bonds,
 the first step must be extract and analyse the frustration network''.
 Given some configuration of ferromagnetic and antiferromagnetic bonds,
 the first step must be to extract and analyze
 the non-frustration network, although
 the effects of frustration are significant,
 since it is considered that
 the local energy on the non-frustration network tends to be lower than 
 that on the frustration network.

 The present bonds are not derived from 
 the Fortuin-Kasteleyn (FK) representation\cite{KF1969, FK1972}
 known as a representation for well studied spin clusters.
 The bonds for the FK representation depend on spin configurations;
 on the other hand, the present bonds do not depend on spin configurations.

\end{document}